\documentclass[twocolumn,preprintnumbers,amsmath,amssymb]{revtex4}

\usepackage{graphicx}
\usepackage[draft]{hyperref}
\usepackage{amsmath}

\newcommand{\hh}{{\bf H}}
\newcommand{\ee}{{\bf E}}

\newcommand{\rmi}{{\rm i}}

\begin{document}

\title{ Can a single gradientless light beam drag particles? }


\author{Andrey Novitsky$^{1}$ and Cheng-Wei Qiu$^{2}$}

\affiliation{$^{1}$DTU Fotonik, Department of Photonics Engineering,
Technical University of Denmark, ${\O}$rsteds~plads~343, DK-2800
Kgs. Lyngby, Denmark, e-mail: anov@fotonik.dtu.dk}

\affiliation{$^{2}$Department of Electrical and Computer
Engineering, National University of Singapore, 4 Engineering Drive
3, Singapore 117576, Singapore. e-mail: eleqc@nus.edu.sg}


\begin{abstract}
Usually a light beam pushes a particle when the photons act upon it.
This is due to that the electric-dipole particle in the paraxial
beam is considered. We investigate the scattering forces in
non-paraxial gradientless beams and find that the forces can drag
certain particles towards the beam source. The major criterion to be
carried out to get the attractive force is the strong
non-paraxiality of the light beam. The cone angle denoting the
non-paraxiality has been investigated to unveil its importance on
achieving dragging force. We hope the attractive forces will be very
useful in nanoparticle manipulation.
\end{abstract}


\maketitle

In spite of the well established theory, experiment and applications
in particle-light interaction
\cite{Ashkin,Chaumet,Nieto,Cizmar,Dholakia}, there are many unknowns
yet to be explored in the particular field of attracting or
separating molecules or nanoparticles by gradientless light beams.
In the ordinary optical tweezers, the transfer of the particles
along the 3D trajectories can be achieved using the spatial light
modulators \cite{Grier}. In this case, the consecutive change of the
computer-generated holograms for the tightly focused light beams
produces the necessary effect, trapping of the particle due to the
gradient of the field.

It is well known that the light beam results in repelling force on
the object. It is correct within electric-dipole approximation and
paraxial light beams, when the size of the particle is much less
than the wavelength \cite{Chaumet}. In this case, the force can
still be manipulated to be a dragging force, provided that two
oppositely directed beams are present to control the particle's
position \cite{Shvedov}. However, such two-beam configuration
(requiring exactly opposite directions) is not necessary to achieve
the attractive force. As shown in Ref. \cite{Sukhov}, two beams with
different longitudinal wavenumbers can locally operate as a tractor
beam. Another possibility to get attractive force is to use the gain
media \cite{Mizrahi}, but the exotic media strongly confine the
possible applications of the \emph{negative} forces.
In Ref. \cite{Novitsky07} it was suggested without proof that the 
negative Poynting vector could be the reason of the attractive force. 
Though the negative energy flux density is not the reason of 
the particle drag towards the light source, this effect is quite 
possible in the non-paraxial beams.

Small particles (hence, the dipole approximation) as a limiting
factor are not necessary either. The particle can be large compared
to the wavelength and positioned near the center of the beam. In
this letter we use a non-paraxial gradientless beam (a vector Bessel
beam) to induce the dragging force. The choice is owing to the
following reasons: (i) long propagation distance, where beams weakly
spread and keep their intensity profile; (ii) complex polarization
structures of the beam; (iii) non-paraxiality of the beams denoted
by the cone angle. We need such a general description of the beam,
because it would be nice to know whether the beam should be actually
complex or we can use its simplified version. Field distribution in
the form of the vector Bessel beam is not unusual. It can be
generated as a mode of a circular fiber. However, the experimental
generation of the non-paraxial beams is still challenging.

The key idea is to demonstrate that a gradientless beam can exert a
dragging force upon a particle. The beam being non-paraxial and
gradientless is very important because the particle can thus be
directly dragged toward the light source realizing the idea of the
tractor beam. It should be noted that the Bessel beam reconstructs
the field behind the particle \cite{Bouchal}. This means that we can
trap and drag several objects by a single beam.

\begin{figure}[tb]
\centerline{\includegraphics[width=8.0cm]{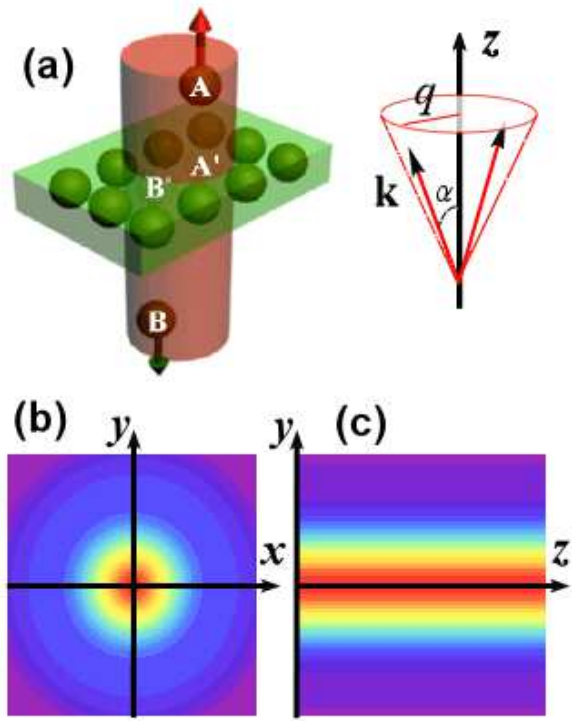}} \caption{ (a)
Light beam pushes a particle from position B' to position B while
drags another particle from original position A' to new position A
(it will keep dragging if the light keep shining). On the right, the
gradientless bessel beam wavevectors lying on the cone are sketched.
(b) and(c) denote the cross-sections of the intensity of such a beam
propagating along z-axis ($c_1 =1$, $c_2 =\rmi$, $q/k_0 = 0.9$, $m =
1$). } \label{fig:1}
\end{figure}

\begin{figure*}[tb]
\centerline{\includegraphics[width=17.0cm]{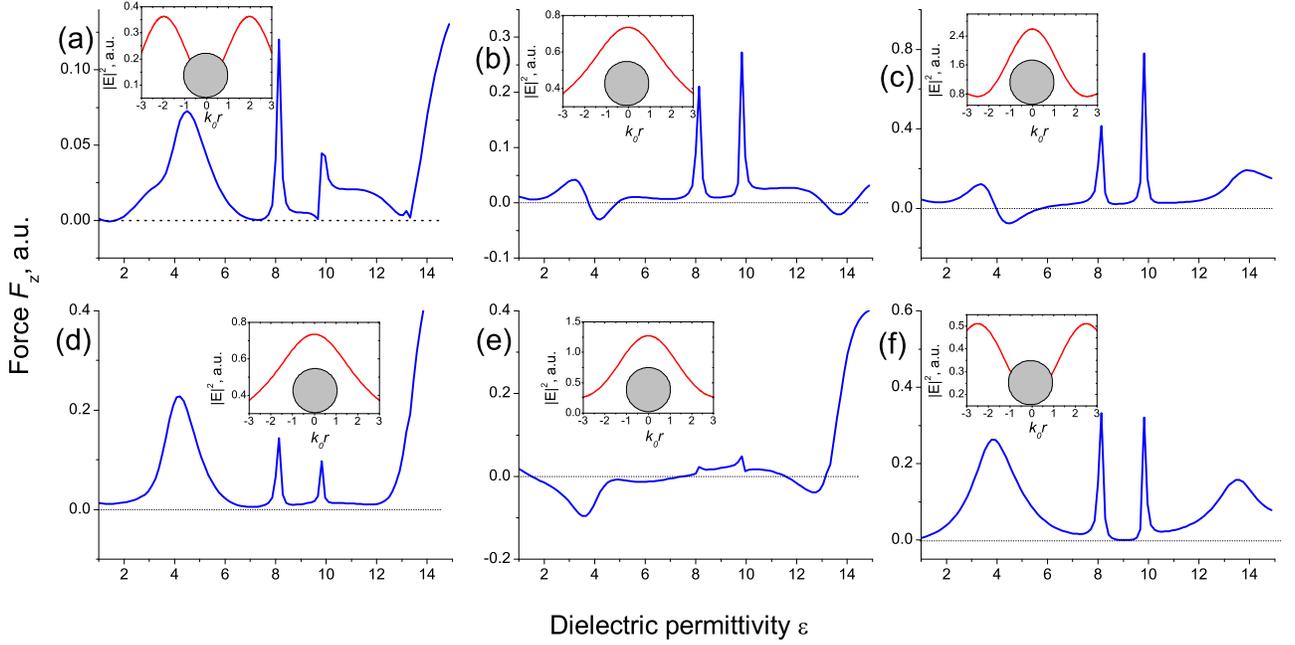}} \caption{
Force $F_z$ on the spherical particle ($k_0 R = 1$, $\mu = 3$)
versus dielectric permittivity $\varepsilon$ for (a) $c_2 = 0$, (b)
$c_2 = 1$, (c) $c_2 = 2$, (d) $c_2 = -1$, (e) $c_2 = \rmi$, and (f)
$c_2 = -\rmi$. In the insets, the field intensities are demonstrated
(grey circle stands for the particle). Beam parameters are $c_1 =1$,
$q/k_0 = 0.9$, and $m = 1$. } \label{fig:2}
\end{figure*}

$z$-propagating electromagnetic Bessel beam in vacuum is described as \cite{Novitsky07}
\begin{eqnarray}
{\bf E}= {\rm e}^{\rmi m \varphi + \rmi \beta z} \left( J_m (q r)
c_2 {\bf e}_z - \frac{k_0}{q} c_1 ({\bf e}_z \times {\bf b})
+ \frac{\beta}{q} c_2 {\bf b} \right), \nonumber \\
{\bf H}= {\rm e}^{\rmi m \varphi + \rmi \beta z} \left( J_m (q r)
c_1 {\bf e}_z + \frac{\beta}{q} c_1 {\bf b} + \frac{k_0}{q} c_2
({\bf e}_z \times {\bf b}) \right),
\label{BB}
\end{eqnarray}
where $k_0 = \omega/c$ is the wavenumber in vacuum, $q=k_0 \sin \alpha$ is the transverse wavenumber
($\alpha$ is the cone angle, see Fig. \ref{fig:1} (a)), $\beta = \sqrt{k_0^2-q^2}$ is
the longitudinal wavenumber, $m$ is the beam order,
\[
{\bf b}=\rmi J'_m (q r)
{\bf e}_r - \frac{m}{q r} J_m (q r) {\bf e}_\varphi, \qquad J'_m (q
r)=\frac{{\rm d} J_m}{{\rm d}(q r)}.
\]
The typical intensity distribution $|\ee|^2$ of the vector Bessel beam is
illustrated in Fig. \ref{fig:1}(b) and (c). The intensity is actually
propagation invariant ($z$-independent) and azimuth invariant ($\varphi$-independent).
It worth to draw attention to the coefficients $c_1$ and $c_2$ of the vector Bessel beam (\ref{BB}).
They provide the possibility to create the appropriately polarized propagation
invariant field distribution and, in particular, can be associated with TE-polarized ($c_2=0$) and
TM-polarized ($c_1=0$) waves. In general, coefficients $c_1$ and $c_2$ can be complex
numbers, what means that the TE and TM beams are phase shifted. The ordinary Bessel
beams follow from Eq. (\ref{BB}) in paraxial approximation ($q \ll k_0$), but then we lose
both possibility to set large $q$s and generate beams
with complex polarization structure. Non-paraxial Bessel beams have already demonstrated
a number of exciting properties like local negative direction of the Poynting vector \cite{Novitsky07} and
intensity transformation on reflection \cite{Novitsky08a}.

We assume that the spherical particle of the radius $R$ is situated
exactly at the center of the beam, but it is not a limitation,
because the calculations can be made for the shifted particle as
well. The incident vector Bessel beam is scattered by the particle
according to the Mie theory. The scattered fields can be calculated,
for example, using the matrix approach \cite{Novitsky08b,Qiu}, which
is valid for any incident electromagnetic wave. So, the fields at
the boundary of the sphere can be written as the sum of the incident
and scattered fields: $\ee = \ee^{inc} + \ee^{sc}$ and $\hh =
\hh^{inc} + \hh^{sc}$. These fields are sufficient to compute the
time-average electromagnetic force on the particle:
\begin{equation}
{\bf F} =  \int_0^\pi \int_0^{2\pi} (\hat{T} {\bf n}) R^2 \sin\theta d\theta d\varphi,
\end{equation}
where ${\bf n}$ is the outward normal, time-averaged Maxwell's stress tensor is
\begin{equation}
\hat{T} = \frac{1}{8\pi} {\rm Re}\left( \ee \otimes \ee^\ast + \hh \otimes \hh^\ast
- \frac{1}{2} (|\ee|^2 + |\hh|^2) \right).
\end{equation}
Here $\ee \otimes \ee^\ast$ defines the dyad, which in index form is
$(\ee \otimes \ee^\ast)_{ij} = E_i E_j^\ast$. We use the general
formulae for calculations, but the scattered fields of the Rayleigh particles
can be written via the electric and magnetic dipole moments \cite{Nieto}.
We are interested only in the component $F_z$ of the force and
look for the situations when $F_z<0$, i.e. when the particle is pulled
by the beam.

When the particle is located at the beam axis, angle $\varphi$
for the Bessel beam (\ref{BB}) is simultaneously the azimuthal angle
of the particle's spherical coordinates. Therefore, the dependence of the fields
on $\varphi$ keeps only as an exponential $\exp(\rmi m \varphi)$ and
the incident and scattered fields contain the single term in the sum over
the integer azimuthal number. It is not the case for the particle shifted
from the axis. Then we need to take into account also the terms with
integer azimuthal numbers nonequal $m$.

The structure of the beam strongly defines the possibility of the
attractive force. However, the intensity of the light is not the
criterium of the pulling. Even identical intensity patterns (like in
the insets of Fig. \ref{fig:2} (b) and (d)) result in the
qualitatively different dependencies of the force. It is likely that
the vector (polarization) structure of the beam plays the leading
role in the appearance of the attractive force. It should be noted
that the negative longitudinal component of the Poynting vector
$S_z$ is not responsible for the appearance of the negative force,
because the large negative $S_z$ exists for $c_2 = -\rmi$
\cite{Novitsky07}, while the force is positive in this case (Fig.
\ref{fig:2}(f)).

Strong attractive force exists for non-phase-shifted superposition
of TE and TM-polarized beams ($c_2$ is positive real number as in
Fig. \ref{fig:2}(b) and (c)) and for $\pi/2$-shifted beams ($c_2$ is
imaginary number with ${\rm Im} c_2>0$ as in Fig. \ref{fig:2}(e)).
In the latter case, the negative force is broadband, low positive
peaks being caused by the quadrupole terms (by the integer polar
number $l=2$ in the scattering series). These two peaks at
$\varepsilon = 8$ and $\varepsilon = 10$ exist in each subfigure and
provide the positive force. The dipole term ($l=1$) describes the
features near $\varepsilon=4$. Due to the non-paraxial regime,
and/or large particle size, and/or magnetic moments the force owing
to the dipole term $l=1$ can become attractive.

\begin{figure}[tb]
\centerline{\includegraphics[width=8.0cm]{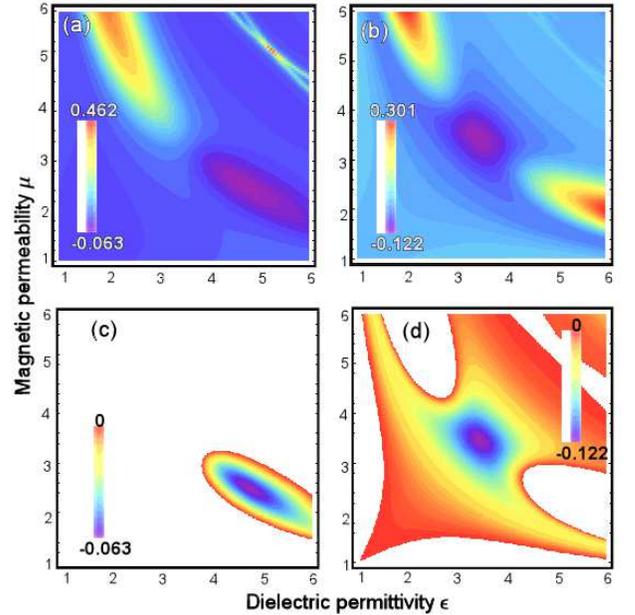}}
\caption{Diagram of the force $F_z$ as function of the dielectric
permittivity $\varepsilon$ and magnetic permeability $\mu$ for (a),
(c) $c_2 = 1$ and (b),(d) $c_2 = \rmi$. In the bottom subfigures,
only negative values of the force is shown. Parameters: $k_0 R = 1$,
$c_1 =1$, $q/k_0 = 0.9$, $m = 1$.} \label{fig:3}
\end{figure}

\begin{figure}[tb]
\centerline{\includegraphics[width=8.0cm]{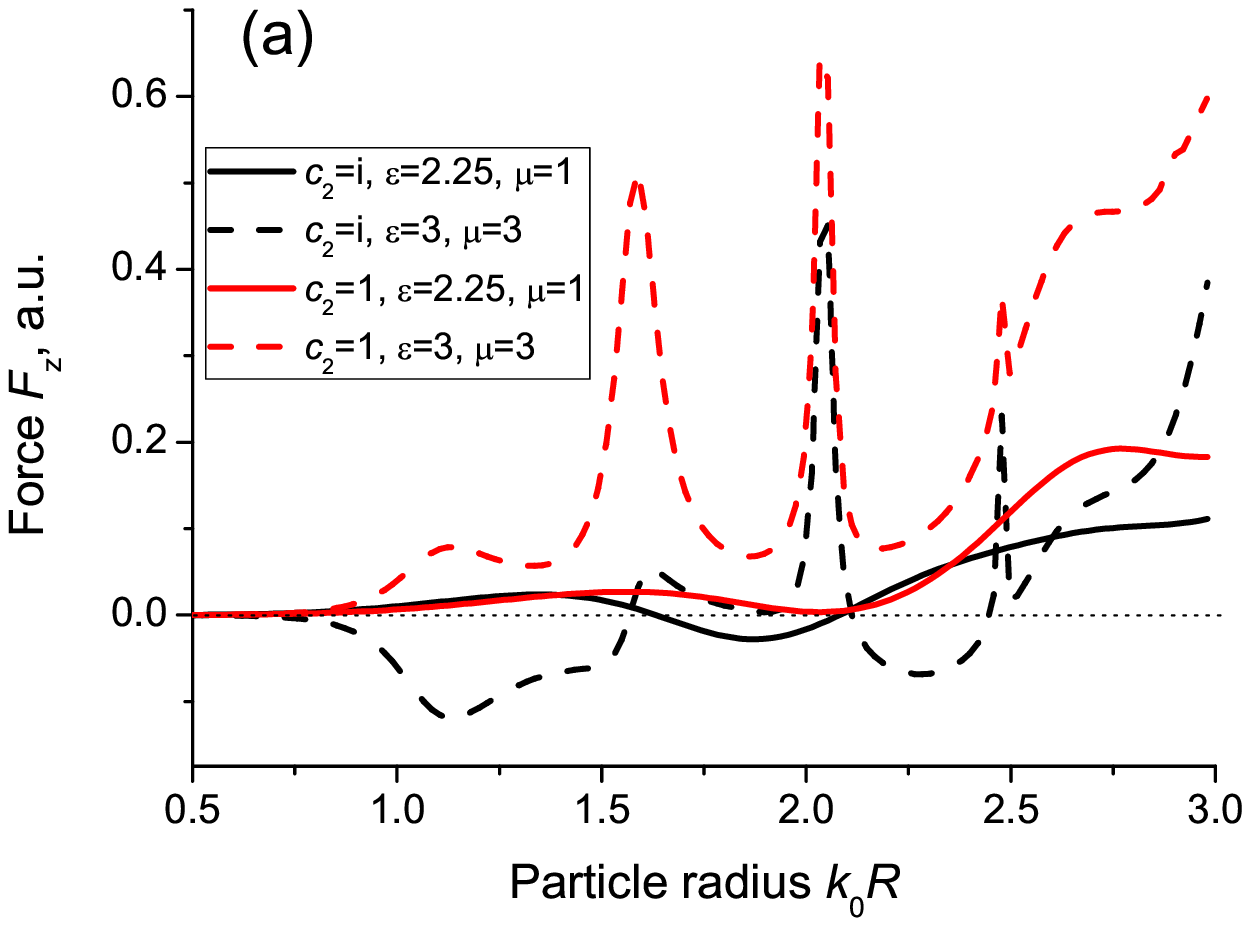}}
\centerline{\includegraphics[width=8.0cm]{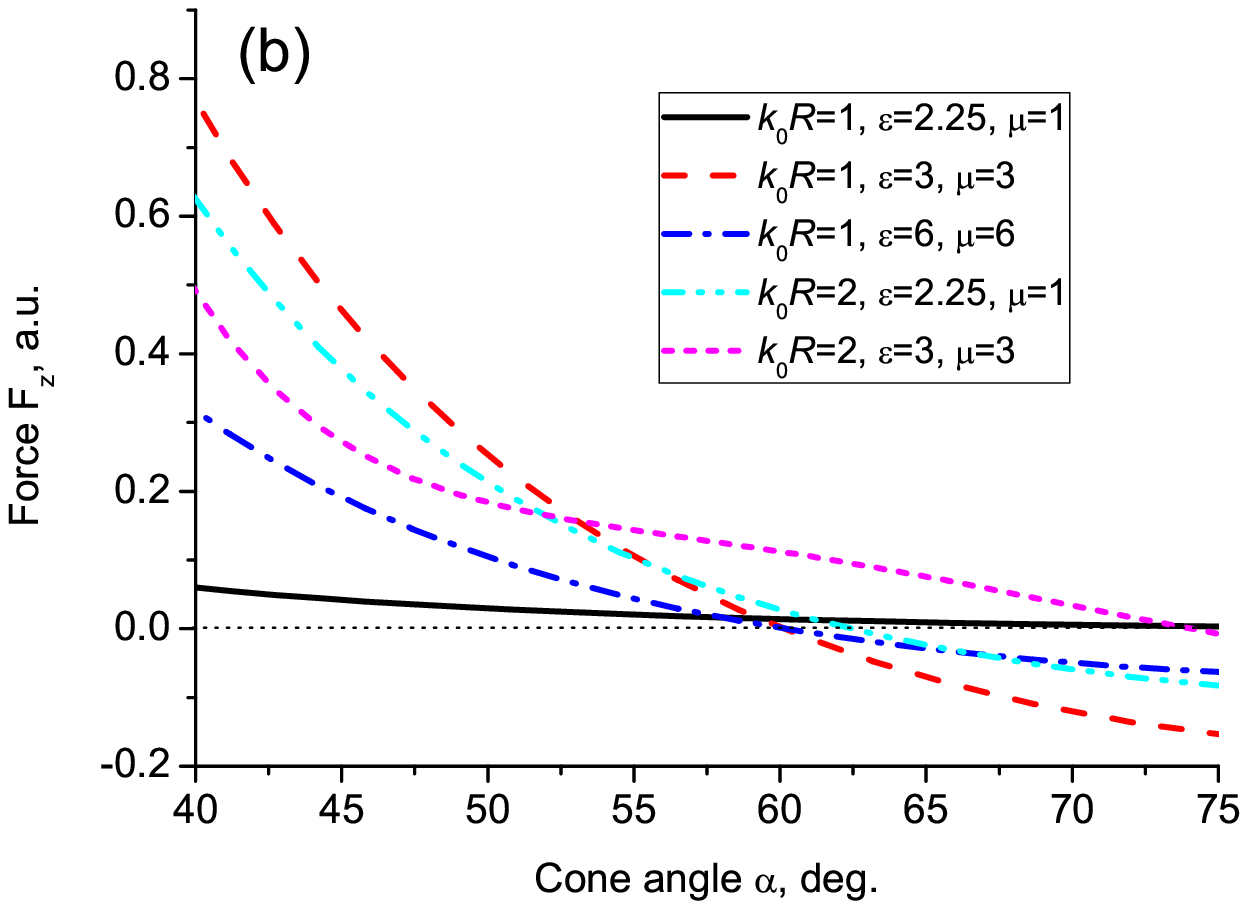}}
\caption{Dependence of the $z$-component of the force on (a) the
radius of the spherical particle $k_0 R$ ($q/k_0 = 0.9$) and (b)
cone angle $\alpha = \arcsin(q/k_0)$ ($c_2 = \rmi$). A multimedia
file is created to show the significance of the cone angle
(non-paraxiality effect) and particle's radius (size effect), in
order to achieve dragging forces (see Media 1. Hyperlink:
http://www.ece.nus.edu.sg/stfpage/eleqc/force.gif). Parameters: $c_1
=1$, $m = 1$.} \label{fig:4}
\end{figure}

Thus, the Bessel beams with $c_2=1$ and $c_2=\rmi$ are preferred,
but they are quite different. In Fig. \ref{fig:3} we show the force
acting on the particle for $c_2=1$ (subfigures (a) and (c)) and
$c_2=\rmi$ (subfigures (b) and (d)). Asymmetry in the case of
$c_2=1$ implies that, in order to induce the attractive force the
permittivity should be greater than the permeability, i.e. the
excited electric moments play the main part. In the symmetric case
(Fig. \ref{fig:3}(b) and (d)), both electric and magnetic inputs are
of the same order and the force has a minimum at the impedance
matching line $\varepsilon = \mu$. It is interesting that the region
of the negative force is really great. The value of the force can be
enhanced by adjusting the radius of the sphere (see Fig.
\ref{fig:4}(a)). From the figure Fig. \ref{fig:3}(d) we can guess
that the force can be attractive even for $\mu = 1$ (it is out of
the range of the plot). It is actually the case as can be seen from
Fig. \ref{fig:4}(a). So, the dragging force can act even on the
common glass particles, if it has an appropriate size. The
maximizing of the absolute value of the negative force using the
impedance matching condition substantially broadens the range of the
negative force. The only strong limitation on the vector Bessel
beams is their non-paraxiality. As shown in Fig. \ref{fig:4} (b),
the variation of the parameters does not substantially reduce the
cone angle $\alpha = \arcsin(q/k_0)$ (for $q=0.9$ the angle $\alpha$
equals $\approx 64^\circ$). The beam should have small longitudinal
wavenumber $\beta$ to be able to create the attractive force.

From Fig. \ref{fig:4}(a) we can suggest that even for the Rayleigh
magnetodielectric particles the force can be less than zero. In
fact, the mechanism behind the attractive force can be revealed from
the theory of magnetoelectric Rayleigh particles \cite{Nieto}. Using
the fact of the small imaginary part of the polarizability in
comparison with the real part, we derive
\begin{eqnarray}
\langle F_z \rangle = \frac{\beta k_0}{2}\left( {\rm Im} (\alpha_e)
|\ee|^2 + {\rm Im} (\alpha_m) |\hh|^2 \right) \nonumber \\
- \frac{k_0^4}{3} {\rm Re} (\alpha_e) {\rm Re} (\alpha_m) {\rm
Re}(P_z), \label{FzMagEl}
\end{eqnarray}
where $P_z = {\bf e}_z (\ee \times \hh^\ast)$, $\alpha_e$ and
$\alpha_m$ are the electric and magnetic polarizabilities of the
particle, respectively. From Eq. (\ref{FzMagEl}) it is evident that
the negative force is feasible due to (i) {\it small} longitudinal
wavenumber $\beta$ of the light beam, (ii) existence of the {\it
magnetic} dipole moment, and (iii) large {\it positive} Poynting
vector (quantity ${\rm Re}(P_z)$).

If $\beta$ is not small, what is the case of the small spherical
particles ($\alpha_e = \alpha_e^{(0)} + \rmi 2 k_0^3 \alpha_e^{(0)
2}/3$, where $\alpha_e^{(0)} = a^3 (\varepsilon - 1)/(\varepsilon +
2)$, and the similar expressions for the magnetic polarizability) in
the field of the plane wave ($\beta = 1$ and $|\ee|^2 = |\hh|^2 =
{\rm Re}(P_z) = 1$) we write
\begin{equation}
\langle F_z \rangle = \frac{k_0^4 a^6}{3}\left(\frac{(\varepsilon -
1)^2}{(\varepsilon + 2)^2} + \frac{(\mu - 1)^2}{(\mu + 2)^2} -
\frac{(\varepsilon - 1)(\mu -1)}{(\varepsilon + 2)(\mu + 2)}
\right). \nonumber
\end{equation}
The force $F_z$ cannot be attractive for any dielectric permittivity
and magnetic permeability under the illumination of a plane wave.

For non-magnetic small particles ($\alpha_m = 0$), the force is
always pushing (it follows straightforwardly from Eq.
(\ref{FzMagEl})). Nevertheless, non-magnetic non-Rayleigh particles
can be attracted owing to the contributions of the higher-order
electrical moments.

Eq. (\ref{FzMagEl}) claims the Poynting vector should have large
positive value. From \cite{Novitsky07} we notice that the largest
value of the Poynting vector is achieved for $c_2 = \rmi$ and this
fact is in full agreement with the calculations in Fig. \ref{fig:2}.
However, we should be careful in generalization of the results for
the Rayleigh particle to the non-Rayleigh ones. Indeed, the Poynting
vector distributions are the same for $c_2 = 1$ and $c_2 = -1$, but
the forces dramatically different in the case $k_0 R = 1$ as it is
observed in Fig. \ref{fig:2} (b) and (d). So, Eq. (\ref{FzMagEl})
does not contradict the existence of the attractive force for
magnetodielectric Rayleigh particles.

In conclusion, we have demonstrated the possibility of the
attractive forces from a gradientless light beam. Compared to the
previously reported tractor beams \cite{Sukhov}, we have the
broadband negative force and do not need two opposite beams with
different wavenumbers \cite{Shvedov} or one gradient beam
\cite{Grier}. The proposed gradientless beam with the proper
manipulation from the non-paraxiality can lead to a lasting dragging
force exerted on particles which can be large or small and made of
any materials.

AN thanks financial support from the Danish Research Council for
Technology and Production Sciences (project THz COW). Authors are
grateful for the stimulating discussions with Prof. C.T. Chan. CWQ
acknowledges the financial support from National University of
Singapore through the Grant R-263-000-574-133.


\end{document}